\documentclass[%
 superscriptaddress,
 amsmath,amssymb,
 aps,
 pra,
 floatfix,
]{revtex4-2}

\usepackage{graphicx}% Include figure files
\usepackage{float}
\usepackage{dcolumn}% Align table columns on decimal point
\usepackage{bm}% bold math
\usepackage[colorlinks,linkcolor=blue,citecolor=blue,urlcolor=blue]{hyperref}% add hypertext capabilities

\usepackage{cases}
\usepackage{soul}%删除线

\usepackage{graphbox}%图片\includegraphics中align=t上对齐
\usepackage{array}
\usepackage{makecell} % 表格内换行

\newcommand{\qket}[1]{\left|{#1}\right\rangle}%|1>右矢
\newcommand{\qbra}[1]{\left\langle{#1}\right|}%<1|左矢
\newcommand{\qproj}[2]{\left.\left\langle{#1}\right|\!{#2}\right\rangle}%<1|2>投影
\usepackage{color}
\definecolor{bluereply}{RGB}{61,112,200}
\definecolor{redreply}{RGB}{220,0,0}
\definecolor{greenreply}{RGB}{0,200,0}
\definecolor{pinkredreply}{RGB}{238,42,104}

\allowdisplaybreaks[2]

\begin{document}

\preprint{APS/123-QED}

\title{Implementation of multiparticle quantum speed limits on observables}

\author{Rui-Heng Miao}
\thanks{R.-H. Miao and Z.-D. Liu contributed equally to this work}
\affiliation{Laboratory of Quantum Information, University of Science and Technology of China, Hefei 230026, China}
\affiliation{Anhui Province Key Laboratory of Quantum Network, University of Science and Technology of China, Hefei 230026, China}
\affiliation{CAS Center For Excellence in Quantum Information and Quantum Physics, University of Science and Technology of China, Hefei 230026, China}
\affiliation{Hefei National Laboratory, University of Science and Technology of China, Hefei 230088, China}

\author{Zhao-Di Liu}
\email{zdliu@ustc.edu.cn}
\affiliation{Laboratory of Quantum Information, University of Science and Technology of China, Hefei 230026, China}
\affiliation{Anhui Province Key Laboratory of Quantum Network, University of Science and Technology of China, Hefei 230026, China}
\affiliation{CAS Center For Excellence in Quantum Information and Quantum Physics, University of Science and Technology of China, Hefei 230026, China}

\author{Chen-Xi Ning}
\affiliation{Laboratory of Quantum Information, University of Science and Technology of China, Hefei 230026, China}
\affiliation{Anhui Province Key Laboratory of Quantum Network, University of Science and Technology of China, Hefei 230026, China}
\affiliation{CAS Center For Excellence in Quantum Information and Quantum Physics, University of Science and Technology of China, Hefei 230026, China}

\author{Yu-Cong Hu}
\affiliation{Laboratory of Quantum Information, University of Science and Technology of China, Hefei 230026, China}
\affiliation{Anhui Province Key Laboratory of Quantum Network, University of Science and Technology of China, Hefei 230026, China}
\affiliation{CAS Center For Excellence in Quantum Information and Quantum Physics, University of Science and Technology of China, Hefei 230026, China}

\author{Hao Zhang}
\affiliation{Laboratory of Quantum Information, University of Science and Technology of China, Hefei 230026, China}
\affiliation{Anhui Province Key Laboratory of Quantum Network, University of Science and Technology of China, Hefei 230026, China}
\affiliation{CAS Center For Excellence in Quantum Information and Quantum Physics, University of Science and Technology of China, Hefei 230026, China}

\author{Chuan-Feng Li}
\email{cfli@ustc.edu.cn}
\affiliation{Laboratory of Quantum Information, University of Science and Technology of China, Hefei 230026, China}
\affiliation{Anhui Province Key Laboratory of Quantum Network, University of Science and Technology of China, Hefei 230026, China}
\affiliation{CAS Center For Excellence in Quantum Information and Quantum Physics, University of Science and Technology of China, Hefei 230026, China}
\affiliation{Hefei National Laboratory, University of Science and Technology of China, Hefei 230088, China}

\author{Guang-Can Guo}
\affiliation{Laboratory of Quantum Information, University of Science and Technology of China, Hefei 230026, China}
\affiliation{Anhui Province Key Laboratory of Quantum Network, University of Science and Technology of China, Hefei 230026, China}
\affiliation{CAS Center For Excellence in Quantum Information and Quantum Physics, University of Science and Technology of China, Hefei 230026, China}
\affiliation{Hefei National Laboratory, University of Science and Technology of China, Hefei 230088, China}

\date{\today}% It is always \today, today,

\begin{abstract}

    The energy-time uncertainty relation limits the maximum speed of quantum system evolution and is crucial for determining whether quantum tasks can be accelerated. However, multiparticle quantum speed limits have not been experimentally explored. In this work, we experimentally verify that both multiparticles and entanglement can accelerate the quantum speed on observables in two-particle systems based on ultrahigh precision control of quantum evolution time. Furthermore, we experimentally prove that the initial quantum state plays a critical role in the quantum speed limits of the entangled systems.
    In addition, we experimentally demonstrate that the upper bound and lower bound of the quantum speed are workable even in a nonunitary Markovian open system with two photons.
    The results obtained based on two-photon experiments have been shown to be generalizable to more particles.
    Our work facilitates the characterization of the dynamic transient properties of complex quantum systems and the control of the quantum speed of large-scale quantum systems.
\end{abstract}

\maketitle

%\tableofcontents

\section{Teaser}

Experiments show that multiparticles, entanglement, and nonunitarity can alter quantum evolution speed limits.

\section{Introduction}

The quantum speed limits provide fundamental bounds on how quickly a quantum state can be transformed. Based on various principles such as energy uncertainty \cite{mandelstam1945}, mean energy \cite{MARGOLUS1998188}, the Bures angle \cite{UHLMANN1992L}, and quantum Fisher information \cite{Taddei2013L}, unified quantum speed limits have been proposed \cite{Levitin2009L, Pires2016L, Shanahan2018L, Okuyama2018L, shuning2021L, Pires2024L, Deffner_2017, doufuquan2023L, Hornedal2023L, Poggi2021L, Ness2022}. The crossover between quantum speed limits has also been experimentally observed \cite{GalNess2021L}. The speed of quantum evolution has both upper and lower bounds \cite{TakahashiNJP2022}. Additionally, factors such as non-Markovian dynamics \cite{Pires2024L, Campo2013, Deffnerprl2013L, Cimmarusti2015L, wuwei2023L, wuwei2022L, Lan_2022L, Funo_2019L}, non-Hermitian effects \cite{Bender2007L, xue2024L}, and entanglement \cite{Giovannetti2003L, Batle2005L, Borras2006L, Frowis2012L, Rudnicki2021L, Pandey2023L, jiyang2024L, Yasmin_2024} can accelerate quantum state transformations. Extensive studies have been conducted on tighter quantum speed limits than previous limits \cite{Campaioli2018L, Campaioli2019L, OConnor2021L, maiziyi2023L}. Based on quantum speed limits, optimal quantum control methods have been developed \cite{Ashhab2012L, Caneva2009L, Hegerfeldt2013L, Brouzos2015L, Aifer_2022L, Basilewitsch2024L}. Furthermore, quantum speed limits have applications in adiabatic quantum computation \cite{Suzuki2020}, quantum thermodynamics \cite{Deffner2010L, DasArpan2021L, VanVu2023L} of heat, chemical work, and entropy \cite{nicholson2020}. Quantum speed limits can also be used in Krylov space \cite{TakahashiPRX2024,TakahashiPRL2025}, topology \cite{VanVu2023LTopological}, ultracold gases \cite{Campo2021L}, superconducting circuits \cite{Yaozu2024}, and quantum batteries \cite{Campaioli2017L, Mohan2021L, Gyhm2024L}. The quantum speed limit has been extended to the classical stochastic processes \cite{TakahashiPRR2023}.

The quantum speed limit has been developed to the evolution of general operators \cite{Hornedal2023L}. Compared with the change rate of the density matrix $\rho$, the quantum speed of the expectation value on observables is more suitable for describing complex quantum systems in practical applications \cite{Fogarty2020, zhangmao2023L, Carabba2022L, MohanBrij2022L, Hornedal2022L}. The expectation value on observable $A$ is given by $a = \langle A \rangle = \text{Tr}\left[\rho A\right]$. The quantum speed, i.e., the change rate of the expectation value $a$, is defined as $\left|\dot{a}\right| = \left|\frac{\text{d}a}{\text{d}t}\right|$. The Mandelstam-Tamm bound \cite{mandelstam1945} is a basic upper bound of the quantum speed and is expressed as follows:
\begin{align}
    \left|\dot{a}\right|=&\left|\frac{\text{d}\langle A \rangle}{\text{d}t}\right|\le 2\Delta A \Delta H.
\end{align}
Here, $\hbar = 1$ and $\Delta A$ and $\Delta H$ represent the standard deviations of the observable $A$ and the Hamiltonian $H$, respectively. This quantum speed limit applies to quantum Fisher information \cite{Yadin2024, Gessner2018}, maximizes the sensitivity in quantum metrology \cite{maleki2023speed, Chu2023, Mirkin2020}, and can be used to determine the calculation frequency in quantum computing \cite{MARGOLUS1998188, McKinney2023, Suzuki2020}. However, this quantum speed limit is relatively loose, providing only an upper bound. Recently, L. P. Garc\'{\i}a-Pintos et al. reported that quantum speed limits on observables were limited by the energy-time uncertainty relation \cite{Luis2022,mySupplementalMaterial}. 
The density matrix $\rho$ can be decomposed into its eigenvalues and eigenstates as $\rho = \sum_j^{2^N} p_j \qket{j}\qbra{j}$. Then the observable $A$ can be expressed in terms of the eigenstates ${\qket{j}}$, where the non-diagonal terms of the matrix represent the coherent component $A_C = \sum_{j\neq k}^{2^N} \qket{j} \qbra{j}A\qket{k} \qbra{k}$, and the diagonal terms represent the incoherent component $A_I = \sum_j^{2^N} \qket{j} \qbra{j}A \qket{j} \qbra{j}$. The standard deviations of the observable $A$ are expressed as $\Delta A_C = \sqrt{\langle A_C^2 \rangle - {\langle A_C \rangle}^2}$, $\Delta A_I = \sqrt{\langle A_I^2 \rangle - {\langle A_I \rangle}^2}$. Next coherent quantum Fisher information and incoherent quantum Fisher information are given by $\mathbb{I}^F_C=2\sum_{j\ne k:p_j+p_k>0}^{2^N}\frac{{\left|\left.\left\langle j \left| \dot{\rho}\right| k \right\rangle \right.\right|}^2}{p_j+p_k}$, $\mathbb{I}^F_I=\sum_{j:p_j>0}^{2^N}\frac{{\left|\left.\left\langle j \left| \dot{\rho}\right| j \right\rangle \right.\right|}^2}{p_j}$. The quantum speed can also be separated into the coherent component $\left|\dot{a}_C\right| = \left|\text{Tr}\left[\dot{\rho} A_C\right]\right|$ and the incoherent component $\left|\dot{a}_I\right| = \left|\text{Tr}\left[\dot{\rho} A_I\right]\right|$, satisfying the relations $\left|\dot{a}_C\right|\le\Delta A_C\sqrt{\mathbb{I}^F_C}$, $\left|\dot{a}_I\right|\le\Delta A_I\sqrt{\mathbb{I}^F_I}$. As the total quantum speed is given by $\left|\dot{a}\right| = \left|\dot{a}_C + \dot{a}_I\right|$, thus, the quantum speed satisfies the following bounds:
\begin{align}
    \max(b_{CI}^{-}, b_{IC}^{-}) \le |\dot{a}| \le \min(b_{CI}^{+}, b_{IC}^{+}).\label{label_eq_upperlower_bound}
\end{align}
Here, $b_{mn}^{(\pm)} = |\dot{a}_m| \pm \Delta A_n \sqrt{{\mathbb{I}^F_n}}$, $m(n)\in\{C,I\}$. The quantum speed limit has both upper and lower bounds that have also been proven in closed quantum systems, based on reference evolution \cite{TakahashiNJP2022}.

\begin{figure}[H]
	\centering
    \includegraphics[width=7in]{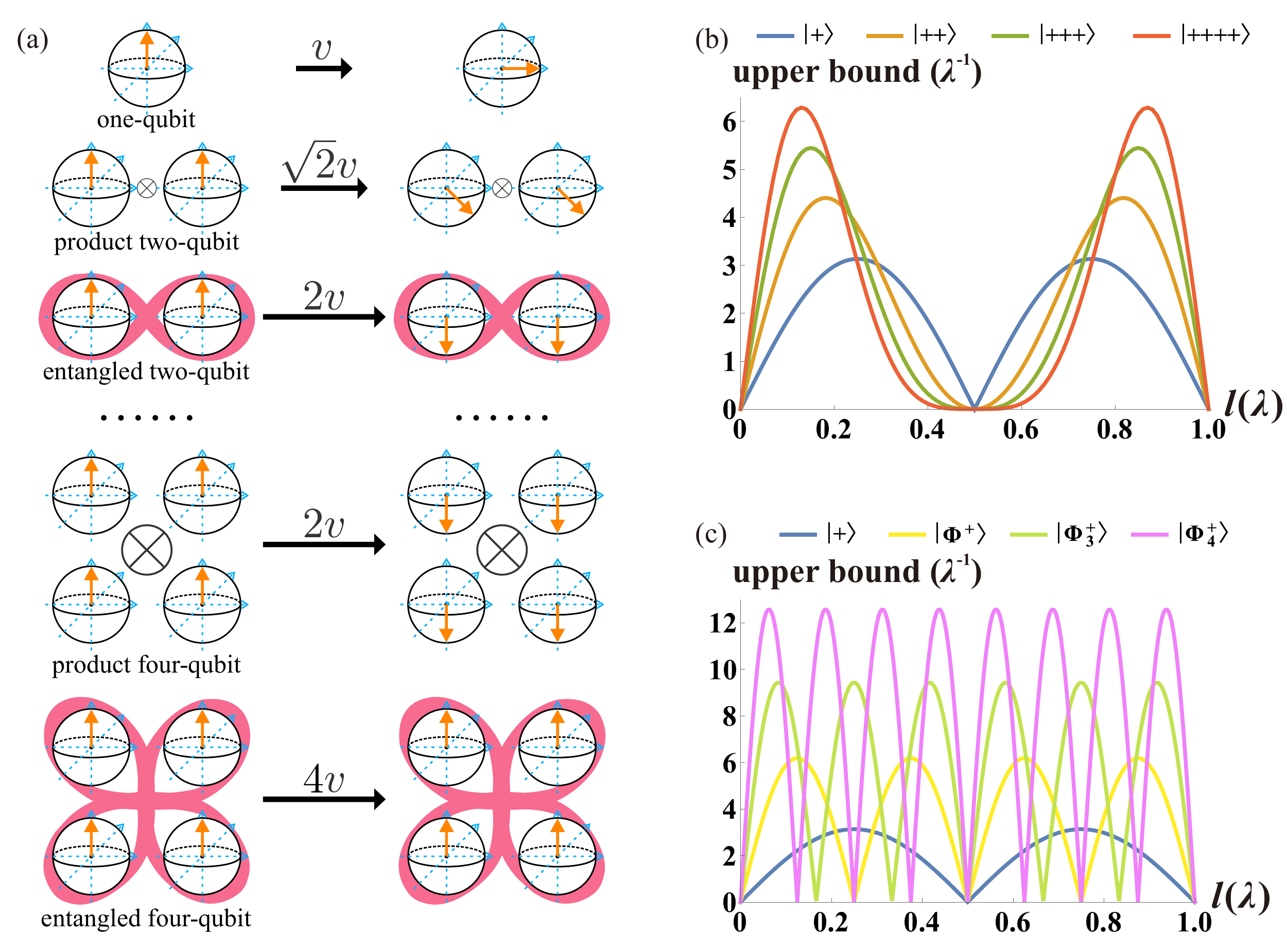}
	\caption{\textbf{Diagram and acceleration effect.} \textbf{(a)} Diagram: multiparticles and entanglement accelerate the quantum speed. A quantum state with faster speed will evolve to a further quantum state simultaneously. \textbf{(b)} Blue, orange, dark green and red lines represent quantum speed limits with the $\qket{+}$, $\qket{++}$, $\qket{+++}$ and $\qket{++++}$ initial states, respectively. Here $\qket{+}=\left(\qket{H}+\qket{V}\right)/\sqrt{2}$. The speedup ratio is $\sqrt{N}$ for product $N$-particle qubit systems, which can be connected with the standard quantum limit. \textbf{(c)} Blue, yellow, light green, and pink lines represent quantum speed limits in frequency correlated systems with the $\qket{+}$, $\qket{\Phi^+}$, $\qket{\Phi_3^+}$, and $\qket{\Phi_4^+}$ initial states, respectively. Here $\qket{\Phi^+}=\left(\qket{HH}+\qket{VV}\right)/\sqrt{2}$, $\qket{\Phi_3^+}=\left(\qket{HHH}+\qket{VVV}\right)/\sqrt{2}$, $\qket{\Phi_4^+}=\left(\qket{HHHH}+\qket{VVVV}\right)/\sqrt{2}$. The speedup ratio is $N$ for entangled $N$-particle qubit systems, which can be connected with the Heisenberg limit. More quantum speed limits numerical results with different four-photon sources and different initial states are presented in Supplemental Material \cite{mySupplementalMaterial}.}
	\label{label_fig_multi_particle}
\end{figure}

Quantum many-body systems often exhibit richer quantum effects \cite{Bukov2019L, Fogarty2020}, which correspond to more valuable applications \cite{Arute2019L, Zheng2024L, liuzd2024L, tang2020low}. F. Yasmin et al. suggested that, in nondegenerate systems, the maximum upper bound of the quantum speed could increase as $1:\sqrt{N}:N$ for single-particle, product $N$-particle, and entangled $N$-particle qubit systems, respectively \cite{Yasmin_2024}, as shown in Fig. \ref{label_fig_multi_particle} (a). Similar conclusions were also obtained by Takahashi et al. in the adiabatic quantum computing \cite{Suzuki2020}. This speedup ratio could be explained by the standard quantum limit and the Heisenberg limit \cite{zhou2018}. In contrast, quantum speed limits contain much richer dynamical details than the Heisenberg limit \cite{Yin2023}. 
However, to the best of our knowledge, multiparticle quantum speed limits remain to be experimentally studied.

We design a proof-of-concept experiment to verify the quantum speed limits on observables using linear optical experiments \cite{liuzd2018L, liuzd2024L, liuzd2020L} in the case of $N=1$ and $N=2$ particles. The quantum speeds of previous experimental works were obtained by fitting a curve from expectation values of a few sparse sampling points, and then deriving the curve to calculate quantum speeds, for example,  by measuring geometric quantum speed limits in a nuclear magnetic resonance experiment\cite{Pires2024L}, by achieving environment-assisted speed-up effect in a cavity quantum electrodynamics experiment \cite{Cimmarusti2015L}, and by observing the quantum speed limits of squeezed states in a superconducting circuit \cite{Yaozu2024}. In this work, we implement a high-density sampling process, which allows us to use the central differences of the expectation values to obtain the quantum speeds directly. The core of realizing a high-density sampling process in linear optical systems is to realize high-precision quantum evolution time control. Using quartz crystal plates, wedge-shaped quartz crystals, and linear motorized stages (LMSs), we achieve a variable-length quartz crystal system. The optical path difference between H polarization and V polarization after passing through the quartz system can be transformed within a wide range with ultrahigh accuracy. 

First, the acceleration effects on the quantum speeds caused by multiparticles and entanglement are confirmed by our experiments. We identify the optimal initial states for single-particle, product two-particle, and entangled two-particle qubit systems, with the speedup ratio of the maximum upper bound as given in \cite{Yasmin_2024}.
Second, we experimentally measure the quantum speed on the observables in a nonunitary Markovian multiparticle system \cite{liuzd2018L,liuzd2024L}. The nonunitary noise will decelerate the maximum quantum speed in multiparticle systems, but can also accelerate the quantum speed in some specific areas. Finally, the upper bound and the lower bound conform to the experimental results whether in multiparticle or nonunitary systems.
Our proof-of-principle experiments highlight the broad application potential of multiparticle quantum speed limits and can be easily extended to frequency correlated multi-photon systems, and frequency decorrelated multi-photon systems \cite{mySupplementalMaterial}, and more platforms such as superconductors and ion traps.

\section{Results}

\subsection{Theoretical Description}

In the $N$-photon dephasing model, we focus on the polarization system interacting with the frequency environment \cite{liuzd2020L, Hamedani2020, liuzd2024L}. The initial total state of the open system is given by:
\begin{align}
    \qket{\psi_{SE}(0)}=&\sum_{i_1,i_2,\cdots,i_N}^{{\{H,V\}}^N}\iint\cdots\int \text{d}\omega_1\text{d}\omega_2\cdots\text{d}\omega_N \Xi(i_1,\omega_1,i_2,\omega_2,\cdots,i_N,\omega_N,0)\qket{i_1,\omega_1}\otimes\qket{i_2,\omega_2}\otimes\cdots\otimes\qket{i_N,\omega_N}.
\end{align}
Here $\Xi(i_1,\omega_1,i_2,\omega_2,\cdots,i_N,\omega_N,0)$ are the coefficients. $\qket{H}$ and $\qket{V}$ are quantum states of the polarization system, and $\qket{\omega_x}$ are quantum states of the frequency environment for the $x$-th particle. The initial total states constructed by various multi-photon sources are shown in Supplemental Material \cite{mySupplementalMaterial}. The initial density matrix of the polarization system is as follows:
\begin{align}
    \rho(0)=&\text{Tr}_E\left[\qket{\psi_{SE}(0)}\qbra{\psi_{SE}(0)}\right]=\iint\cdots\int \text{d}\omega_1\text{d}\omega_2\cdots\text{d}\omega_N \qproj{\omega_1,\omega_2,\cdots,\omega_N}{\psi_{SE}(0)}\qproj{\psi_{SE}(0)}{\omega_1,\omega_2,\cdots,\omega_N}.
\end{align}
Here, $\rho(0)$ is a $N$-qubit density matrix, $\text{Tr}_E$ means to clear the frequency environment information and only keep the information of the polarization system. Note that $\rho(0)$ can be pure even in multiparticle systems, so we can rewrite $\rho(0)$ as $\rho(0)=\qket{\psi(0)}\qbra{\psi(0)}$.

After passing through a birefringent quartz crystal with an optic axis aligned in the $\qket{H}$ direction with a length $L$, the system undergoes additional phase shifts due to the polarization-dependent and frequency-dependent refractive indices. An extra phase $\exp\left[\text{i}n_{H/V}(\omega)\omega L/c\right]$ is introduced for the $\omega$ frequency and H/V polarization along each path. Here, $n_{H/V}(\omega)$ represents the refractive indices of the H/V polarization. To simplify the evolution equation, we define the optical path difference between H polarization and V polarization as $l=(\bar{n}_H-\bar{n}_V)L$; this is used in our experiments to express the crystal length instead of $L$. Here, $\bar{n}_H$ and $\bar{n}_V$ are the average refractive indices of the H polarization (extraordinary rays) and V polarization (ordinary rays), respectively. Throughout this paper, we express the optical path difference $l$ in units of $\lambda$. Here, $\lambda$ is the center wavelength. Therefore, the final total state of the open system is as follows \cite{mySupplementalMaterial}:
\begin{align}
    \qket{\psi_{SE}(l)}=&\sum_{i_1,i_2,\cdots,i_N}^{{\{H,V\}}^N}\iint\cdots\int \text{d}\omega_1\text{d}\omega_2\cdots\text{d}\omega_N \Xi(i_1,\omega_1,i_2,\omega_2,\cdots,i_N,\omega_N,l)\qket{i_1,\omega_1}\otimes\qket{i_2,\omega_2}\otimes\cdots\otimes\qket{i_N,\omega_N}.
\end{align}
Here $\Xi(i_1,\omega_1,i_2,\omega_2,\cdots,i_N,\omega_N,l)=\exp\left[\text{i}\frac{n_{i_1}(\omega_1)\omega_1+n_{i_2}(\omega_2)\omega_2+\cdots+n_{i_N}(\omega_N)\omega_N}{(\bar{n}_H-\bar{n}_V)c}l\right]\Xi(i_1,\omega_1,i_2,\omega_2,\cdots,i_N,\omega_N,0)$.

If we regard the optical path difference $l$ as the evolution time, the total Hamiltonian of the entire open system, including the polarization system and the frequency environment, is as follows:
\begin{align}
    H_\text{SE}=& H_0 \otimes I_{SE}^{\otimes (N-1)} + I_{SE} \otimes H_0 \otimes I_{SE}^{\otimes (N-2)} + \cdots + I_{SE}^{\otimes (N-1)} \otimes H_0.
\end{align}
Here $H_0=\int \text{d}\omega \left[ \frac{n_H(\omega)\omega}{(\bar{n}_H-\bar{n}_V)c}\qket{H,\omega}\qbra{H,\omega} + \frac{n_V(\omega)\omega}{(\bar{n}_H-\bar{n}_V)c}\qket{V,\omega}\qbra{V,\omega} \right]$ is the Hamiltonian on single-particle open system. $I_{SE}^{\otimes N}$ is the direct product of $N$ identity matrices, for example, $I_{SE}^{\otimes 2}=I_{SE}\otimes I_{SE}$. $I_{SE}$ is the identity matrix of single-particle open systems.

For a specific $l$, the density matrix of the polarization system is the trace of the frequency environment of the final total state as follows:
\begin{align}
    \rho(l)=&\text{Tr}_E\left[\qket{\psi_{SE}(l)}\qbra{\psi_{SE}(l)}\right].
\end{align}

The speedup ratio in our system can be directly obtained. The quantum speed with a mixed state must be less than or equal to that in a pure state; thus, the maximum quantum speed occurs in pure states, which can only be achieved in a single frequency environment $\iint\cdots\int \text{d}\omega_1\text{d}\omega_2\cdots\text{d}\omega_N \delta(\omega_1-\omega)\delta(\omega_1-\bar{\omega})\delta(\omega_2-\bar{\omega})\cdots\delta(\omega_N-\bar{\omega})\qket{\omega_1,\omega_2,\cdots,\omega_N}$. Here, $\delta$ is the Dirac delta function, $\bar{\omega}$ is the center frequency. Now, the Hamiltonian of the polarization system with a single frequency environment is \cite{mySupplementalMaterial}:
\begin{align}
    H=&\frac{\pi}{\lambda}\left[\sigma_z\otimes I^{\otimes (N-1)} + I\otimes \sigma_z\otimes I^{\otimes (N-2)} + \cdots + I^{\otimes (N-1)} \otimes \sigma_z\right].\label{label_eq_H_system}
\end{align}
Here, $\sigma_z$ is a Pauli matrix, $I$ is the identity matrix in single-particle polarization systems. Under the above conditions, the upper bound in Eq. \eqref{label_eq_upperlower_bound} simplifies to the following \cite{mySupplementalMaterial}:
\begin{align}
    \left|\dot{a}\right|\le&\sqrt{a}\sqrt{1-a}\sqrt{\mathbb{I}^F_C}.
\end{align}
The term $\sqrt{a}\sqrt{1-a}$ reaches its maximum value, $1/2$, when $a=1/2$, which is achievable in almost all scenarios by choosing a suitable observable. The coherent quantum Fisher information $\sqrt{\mathbb{I}^F_C}$ remains constant as $l$ changes and depends only on the initial state \cite{mySupplementalMaterial}. The optimal initial state in the single-particle qubit system is $\qket{+}$. The states $\qket{\pm}$ are defined as follows:
\begin{align}
    \qket{\pm}=&\left(\qket{H}\pm\qket{V}\right)/\sqrt{2}.
\end{align}
The optimal initial state in the product $N$-particle system is a direct product of $N$ states of $\qket{+}$: $\qket{+}^{\otimes N}=\qket{+}\otimes\qket{+}\otimes\cdots\otimes\qket{+}$. For the entangled $N$-particle system, the optimal initial state is $\left(\qket{H}^{\otimes N} + \qket{V}^{\otimes N}\right)/\sqrt{2}$. The maximum quantum speed limits for the above three quantum states are $\pi$, $\sqrt{N}\pi$, and $N\pi$  \cite{mySupplementalMaterial}.

In our experiment, we use $\qket{+}$, $\qket{++}$, and $\qket{\Phi^+}$ to demonstrate the maximum quantum speed limits. Here, $\qket{\Phi^+}$ is one of the Bell states shown as follows:
\begin{align}
    \qket{\Phi^\pm}=&\left(\qket{HH}\pm\qket{VV}\right)/\sqrt{2}.
\end{align}
As shown in Fig. \ref{label_fig_multi_particle} (b, c), the blue, orange, and yellow lines represent the quantum speed limits for these initial states in frequency correlated systems. The maximum upper bounds of the quantum speed align with our theoretical predictions. 

To build a nonunitary two-particle system, additional noise for each particle is needed after the previous steps. We introduce a long birefringent crystal, whose length is approximately the coherence length of the system, as $\sigma_x$ noise, whose Hamiltonian of the polarization system is obtained by replacing $\sigma_z$ in Eq. \eqref{label_eq_H_system} with $\sigma_x$, and whose evolution time is a fixed constant; here, $\sigma_x$ is a Pauli matrix. To enhance the noise, we select an initial state far from $\qket{+}$. As given in \cite{Luis2022}, the special state $\qket{P}$ is defined as $\qket{P}\qbra{P}=\frac{1}{2}\left(I+\frac{1}{\sqrt{3}}\sigma_x+\frac{1}{\sqrt{3}}\sigma_y+\frac{1}{\sqrt{3}}\sigma_z\right)$; here, $\sigma_y$ is also a Pauli matrix \cite{mySupplementalMaterial}; then, the state $\qket{P}$ can be rewritten as follows:
\begin{align}
    \qket{P}=&\frac{\left(1-\text{i}\right)\left(1+\sqrt{3}\right)}{2\sqrt{3+\sqrt{3}}}\qket{H}+\frac{1}{\sqrt{3+\sqrt{3}}}\qket{V}.
\end{align}
We then extend this special state to $N$-particle systems as $\qket{P}^{\otimes N}$. For the two-particle optical system, we use the state $\qket{PP}$ as the initial state.
Although the noisy state is no longer pure, we can still set an observable of the initial state $A = \qket{PP}\qbra{PP}$. Additionally, nonunitary noises can break the tightness of the lower bound \cite{Luis2022,mySupplementalMaterial}. If we don't consider mixed states, nonunitary noises and complex environments during the complete evolution process, the lower bound of quantum speed limits can be tight. On the contrary, if we introduce nonunitary noises, the quantum speed departs from the lower bound and approaches the upper bound. Moreover, although the $\sigma_x$ noise can slow the maximum quantum speed in multiparticle systems to the level of single-particle systems, the quantum speed in some specific areas can also be accelerated.

\subsection{Experimental Results}

The two-photon process needed in the experiment is prepared in two layers of nonlinear crystals. The environment is the frequency degree of freedom of the photons, and the open quantum system is the polarization degree of freedom of the photons. The dephasing model is realized by birefringent crystals. Here, the birefringent crystals are a series of quartz crystals \cite{liuzd2024L,liuzd2018L,liuzd2020L}. Ultrahigh precision and wide-range control of the length of the quartz crystals are achieved using movable wedge-shaped quartz crystals \cite{mySupplementalMaterial}. Finally, the tomography process can fully extract the polarization information within the open system. The experimental setup is presented in Fig. \ref{label_fig_setup}, and further details are given in the "Experimental Design" section.

\begin{figure}[H]
    \begin{minipage}[t][1in][t]{0.3in}\vspace{0em}(a)\end{minipage}\hspace{-0.1in}\includegraphics[width=1.5in,align=t]{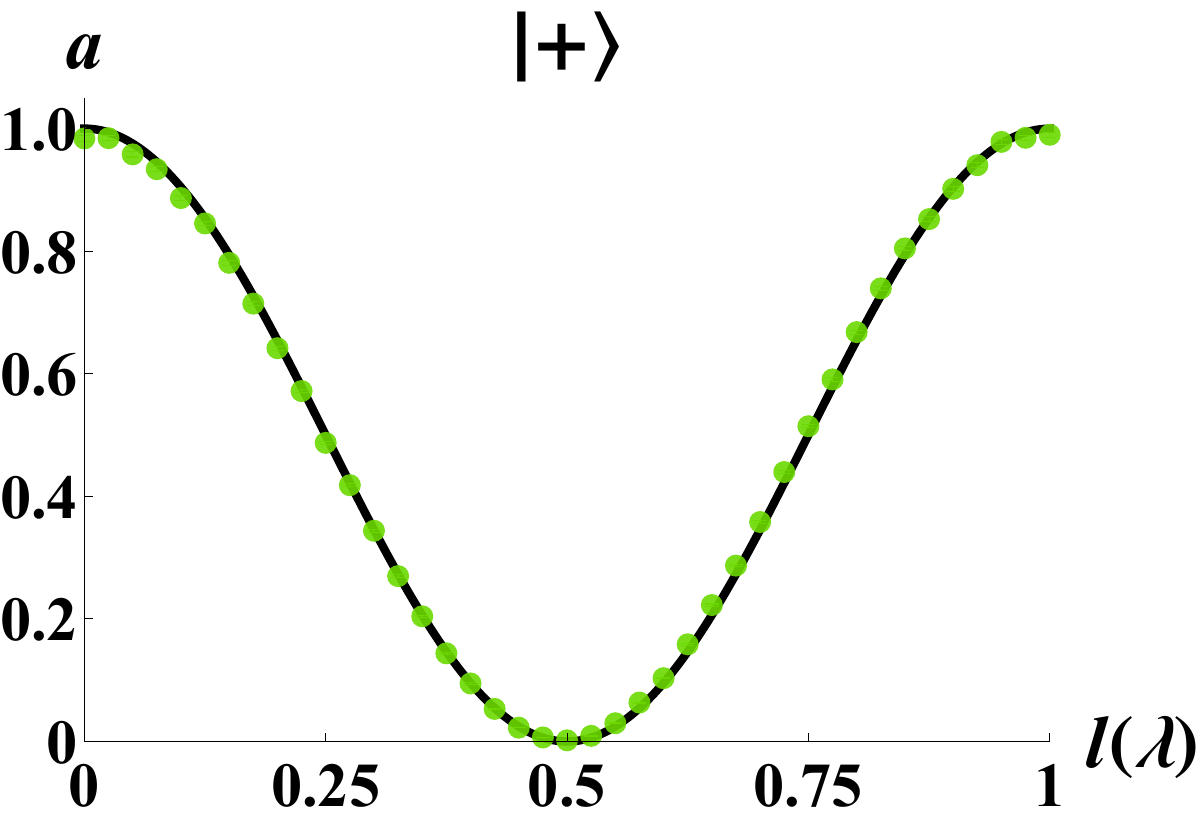}
    \begin{minipage}[t][1in][t]{0.3in}\vspace{0em}(b)\end{minipage}\hspace{-0.1in}\includegraphics[width=1.5in,align=t]{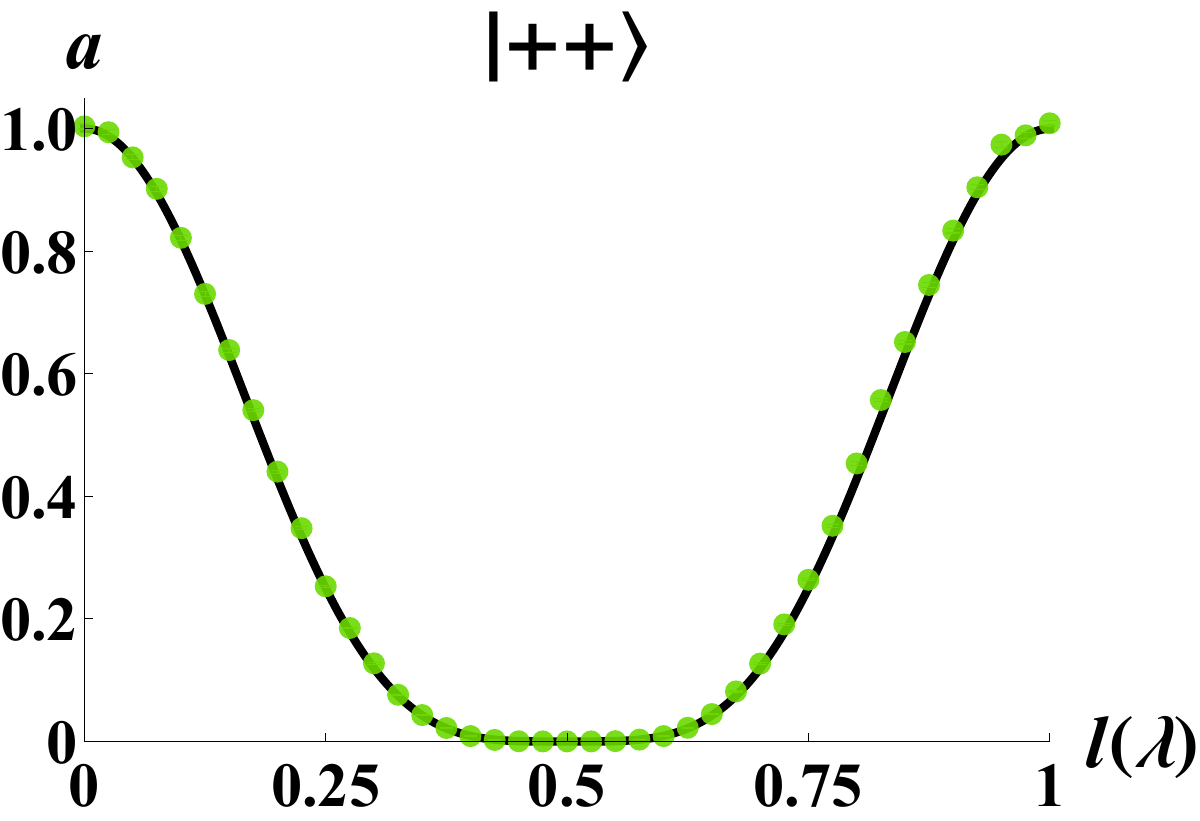}
    \begin{minipage}[t][1in][t]{0.3in}\vspace{0em}(c)\end{minipage}\hspace{-0.1in}\includegraphics[width=1.5in,align=t]{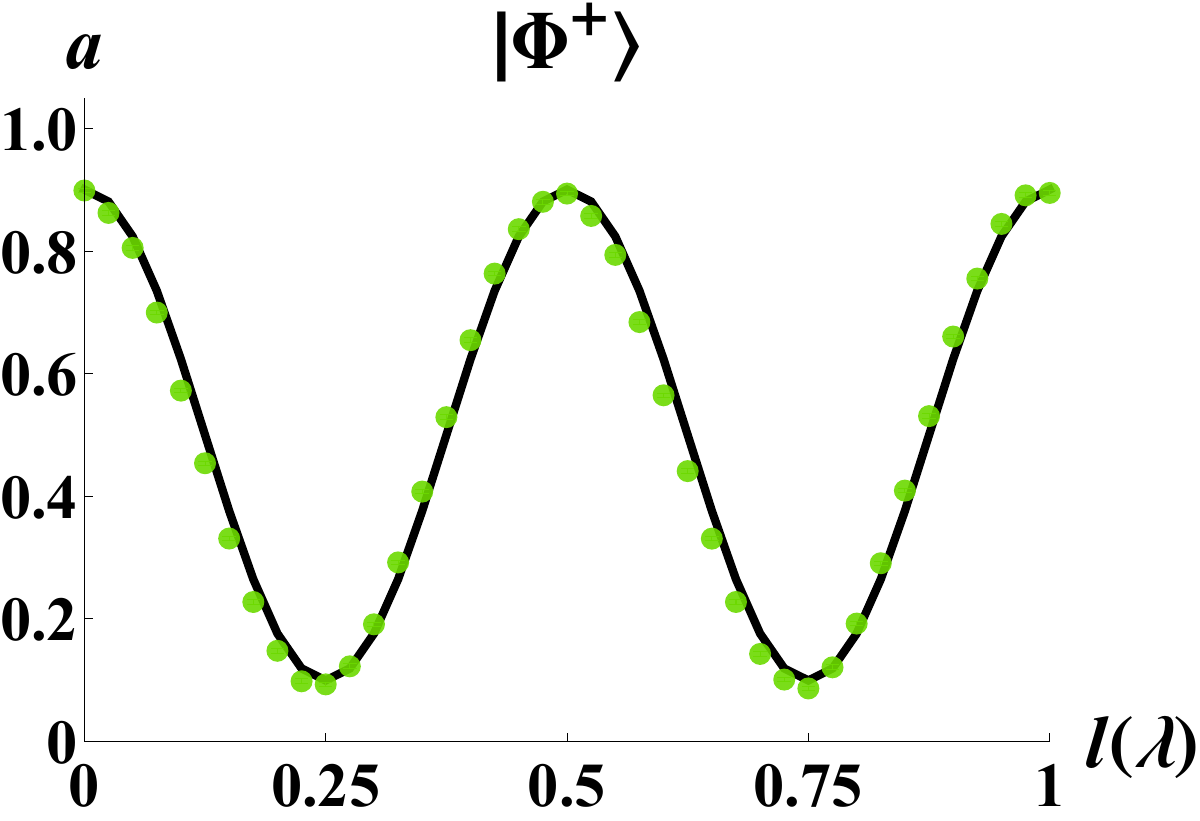}
    \begin{minipage}[t][1in][t]{0.3in}\vspace{0em}(d)\end{minipage}\hspace{-0.1in}\begin{minipage}[t][1in][t]{0.3in}\vspace{0em}
    \begin{center}
    \begin{tabular}{|c|c|}
        \hline
        subfigure & \makecell[c]{maximum \\ quantum \\ speed ($\lambda^{-1}$)}\\\hline
        (e) & $3.103\pm0.006$ \\\hline
        (f) & $4.049\pm0.012$ \\\hline
        (g) & $4.981\pm0.011$ \\\hline
        (l) & $2.184\pm0.004$ \\\hline
        (m) & $3.296\pm0.010$ \\\hline
        (n) & $1.448\pm0.004$ \\\hline
        (o) & $1.391\pm0.004$ \\\hline
    \end{tabular}
    \end{center}
    \end{minipage}
    \vspace{0.1in}\\
    \begin{minipage}[t][1in][t]{0.3in}\vspace{0em}(e)\end{minipage}\hspace{-0.1in}\includegraphics[width=1.5in,align=t]{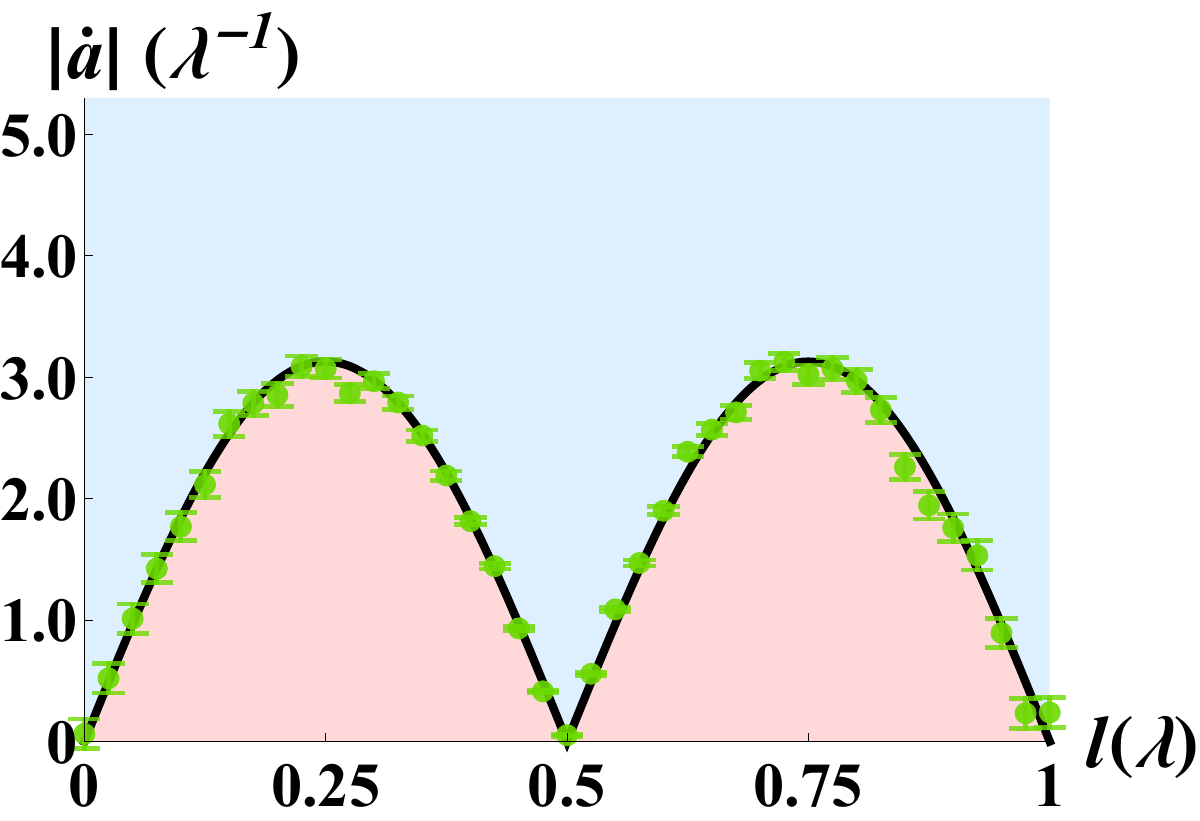}
    \begin{minipage}[t][1in][t]{0.3in}\vspace{0em}(f)\end{minipage}\hspace{-0.1in}\includegraphics[width=1.5in,align=t]{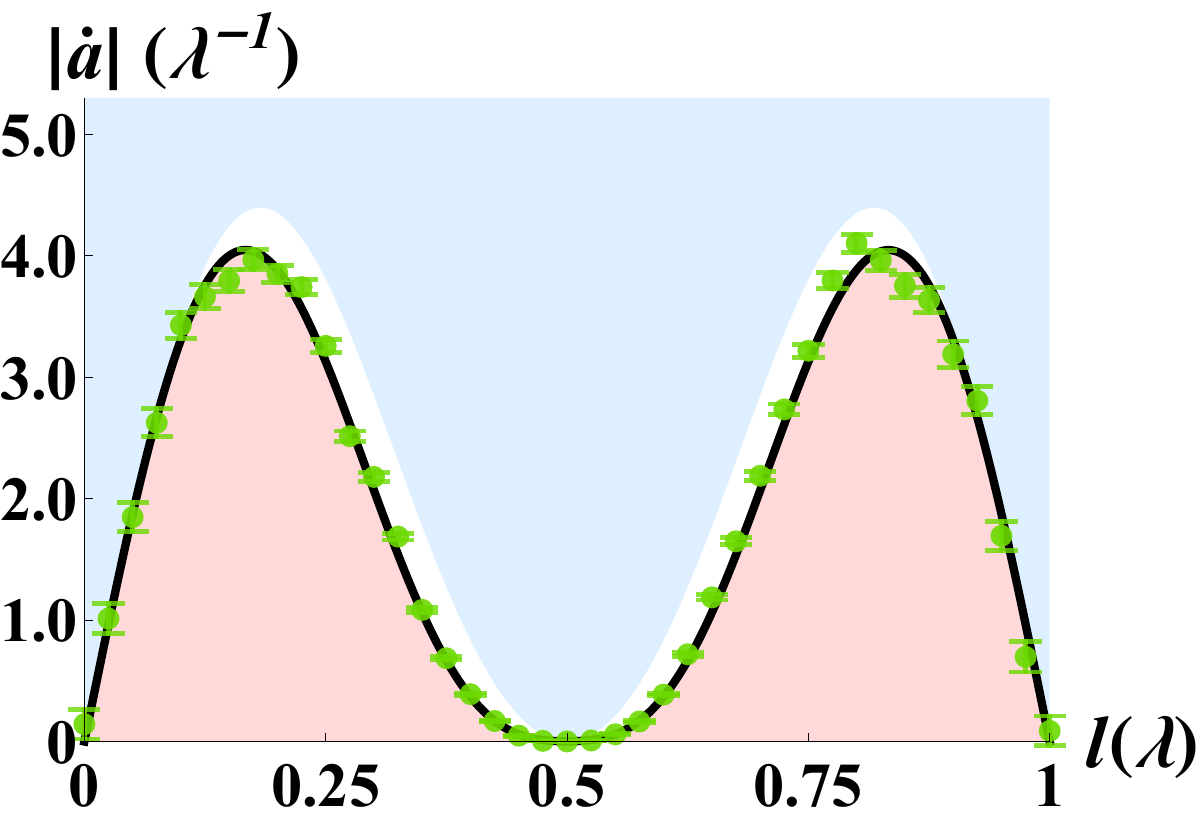}
    \begin{minipage}[t][1in][t]{0.3in}\vspace{0em}(g)\end{minipage}\hspace{-0.1in}\includegraphics[width=1.5in,align=t]{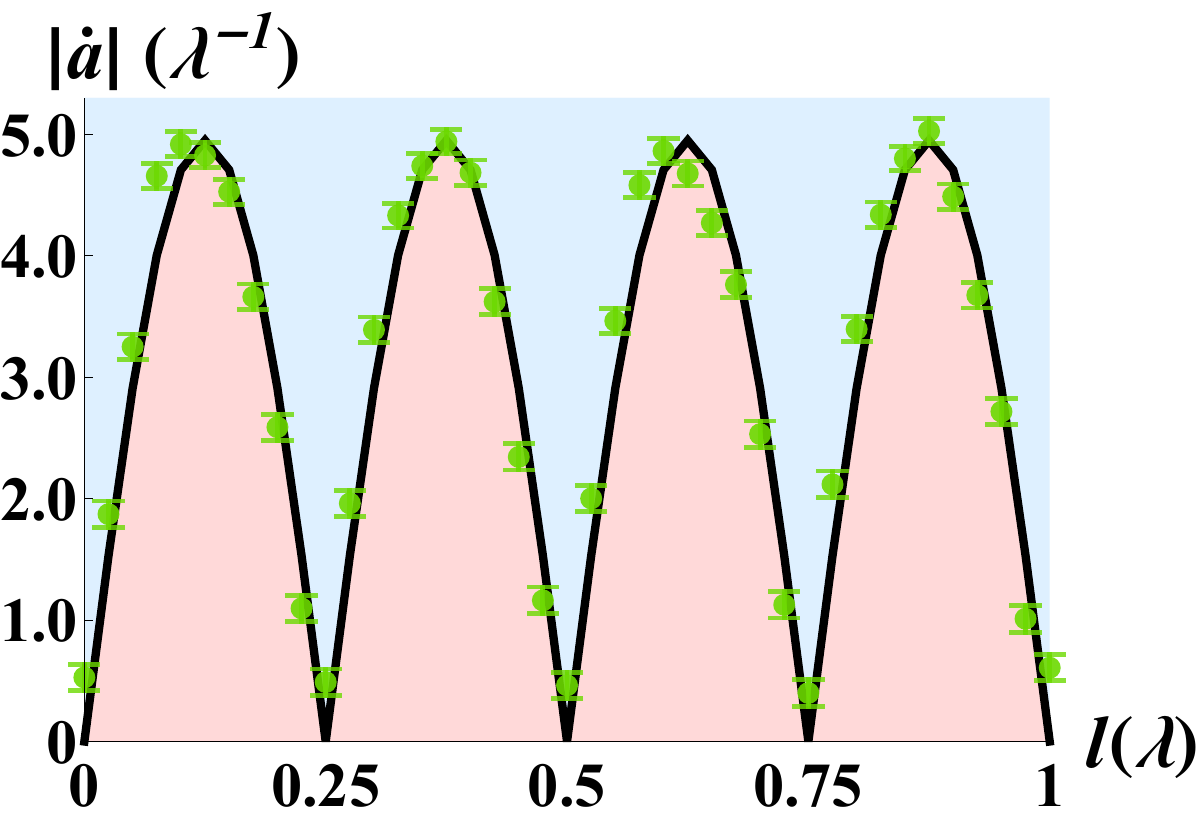}
    \vspace{0.1in}\\
    \begin{minipage}[t][1in][t]{0.3in}\vspace{0em}(h)\end{minipage}\hspace{-0.1in}\includegraphics[width=1.5in,align=t]{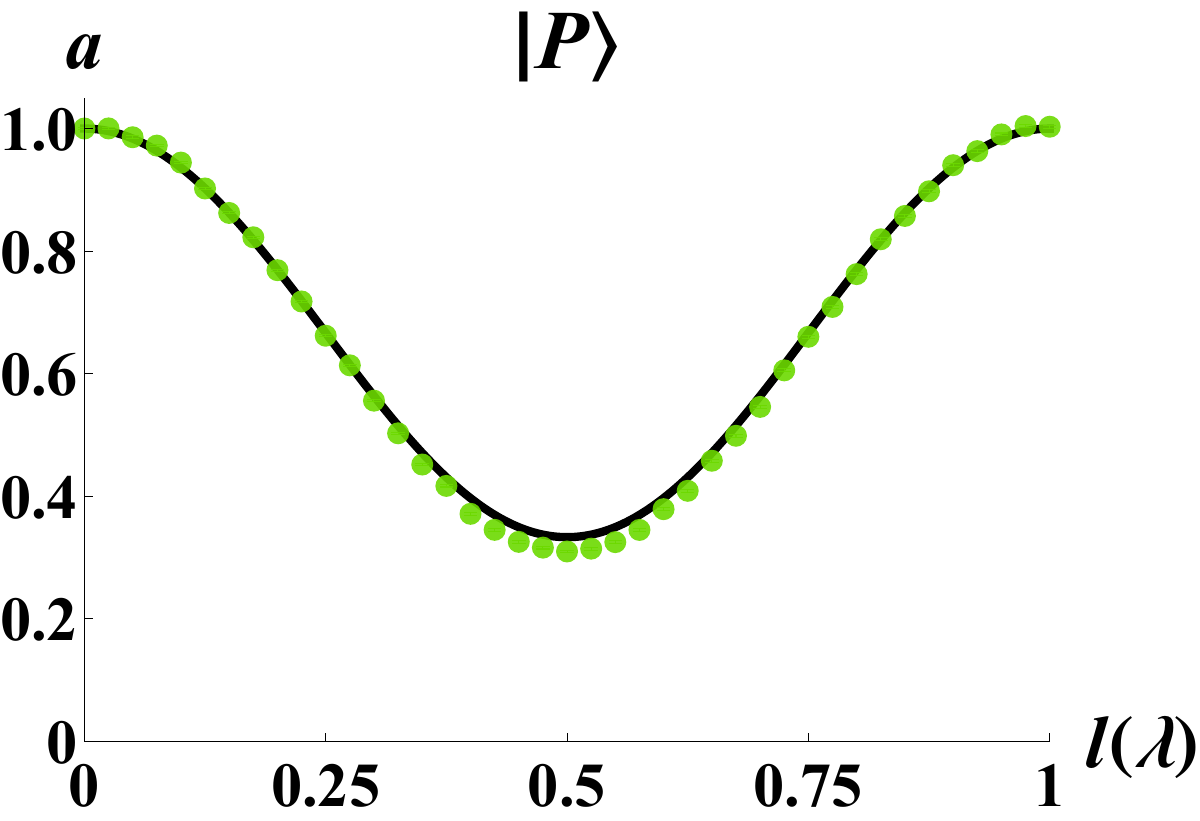}
    \begin{minipage}[t][1in][t]{0.3in}\vspace{0em}(i)\end{minipage}\hspace{-0.1in}\includegraphics[width=1.5in,align=t]{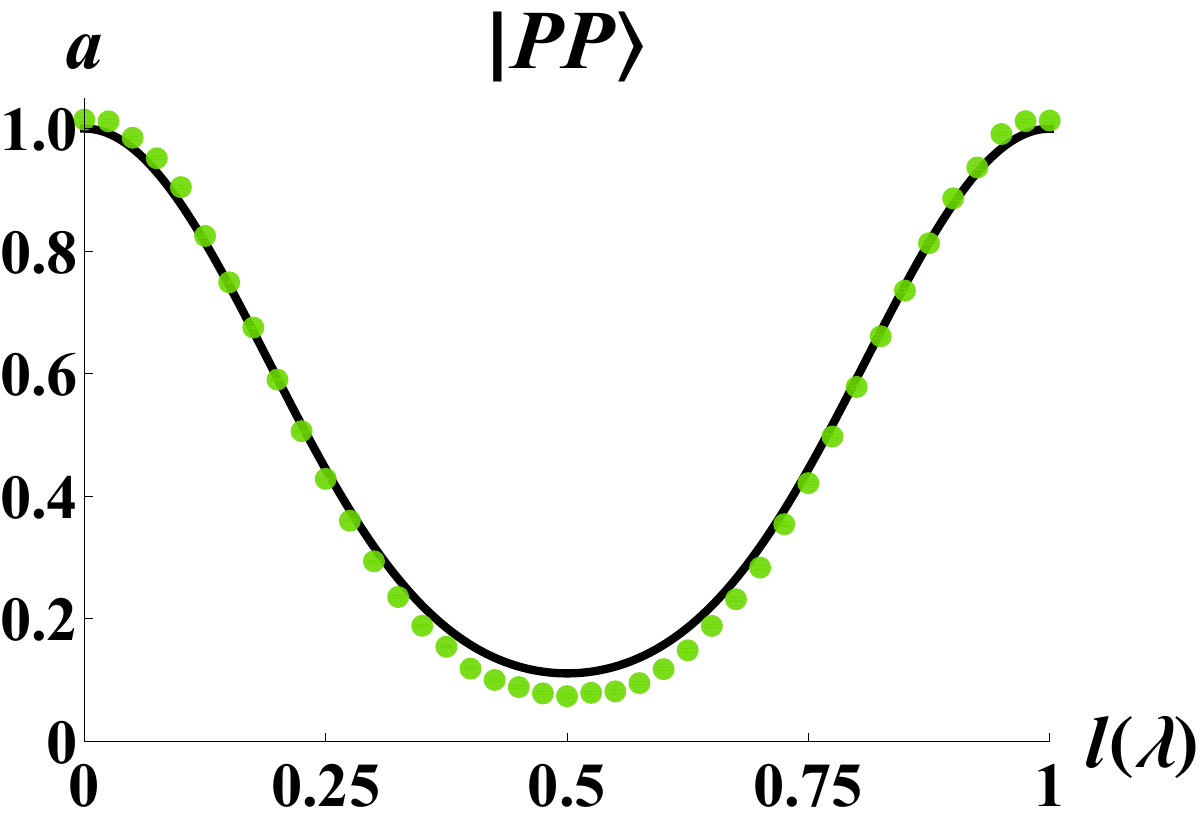}
    \begin{minipage}[t][1in][t]{0.3in}\vspace{0em}(j)\end{minipage}\hspace{-0.1in}\includegraphics[width=1.5in,align=t]{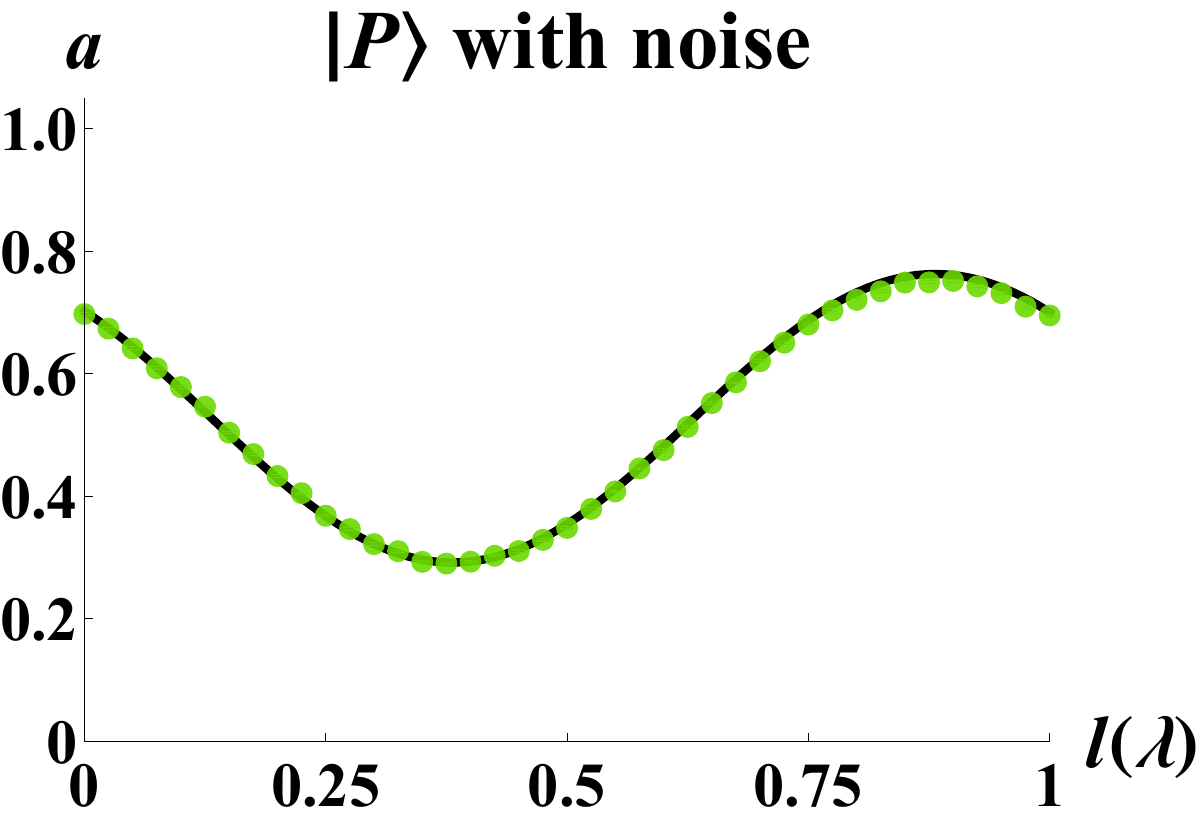}
    \begin{minipage}[t][1in][t]{0.3in}\vspace{0em}(k)\end{minipage}\hspace{-0.1in}\includegraphics[width=1.5in,align=t]{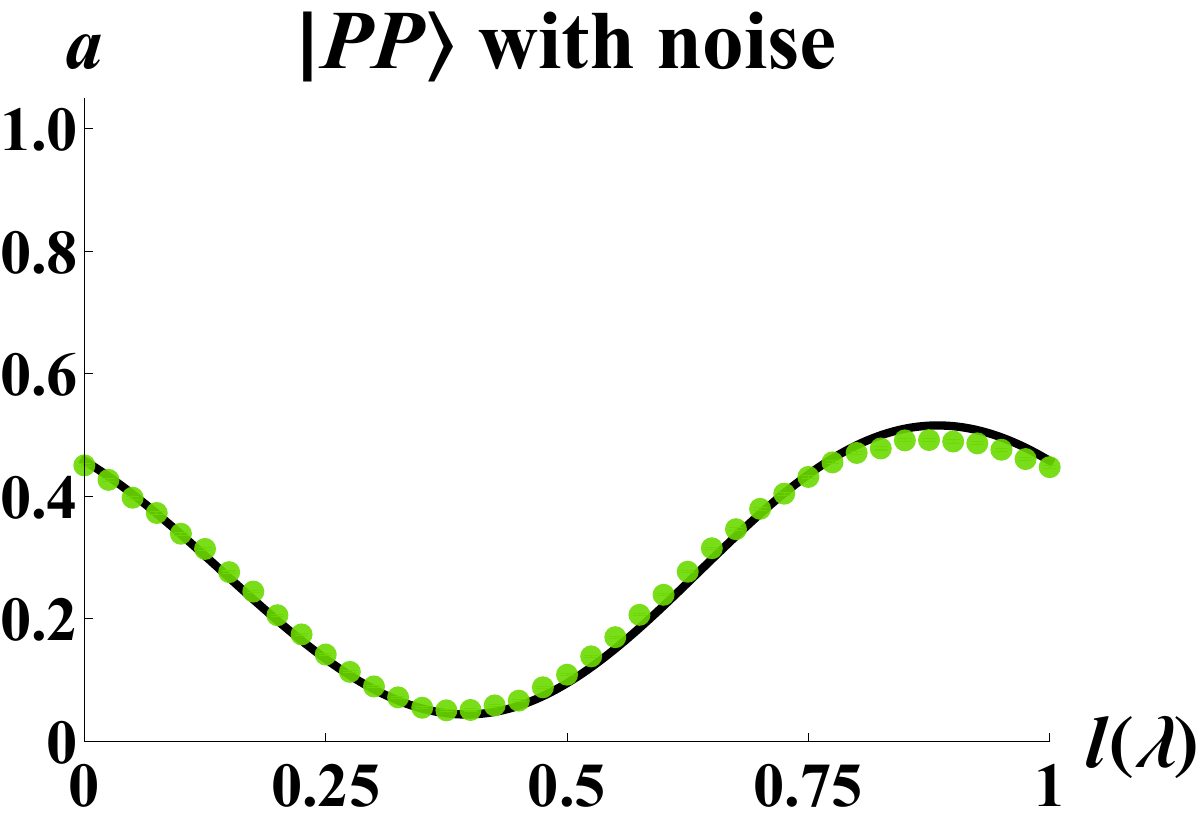}
    \vspace{0.1in}\\
    \begin{minipage}[t][1in][t]{0.3in}\vspace{0em}(l)\end{minipage}\hspace{-0.1in}\includegraphics[width=1.5in,align=t]{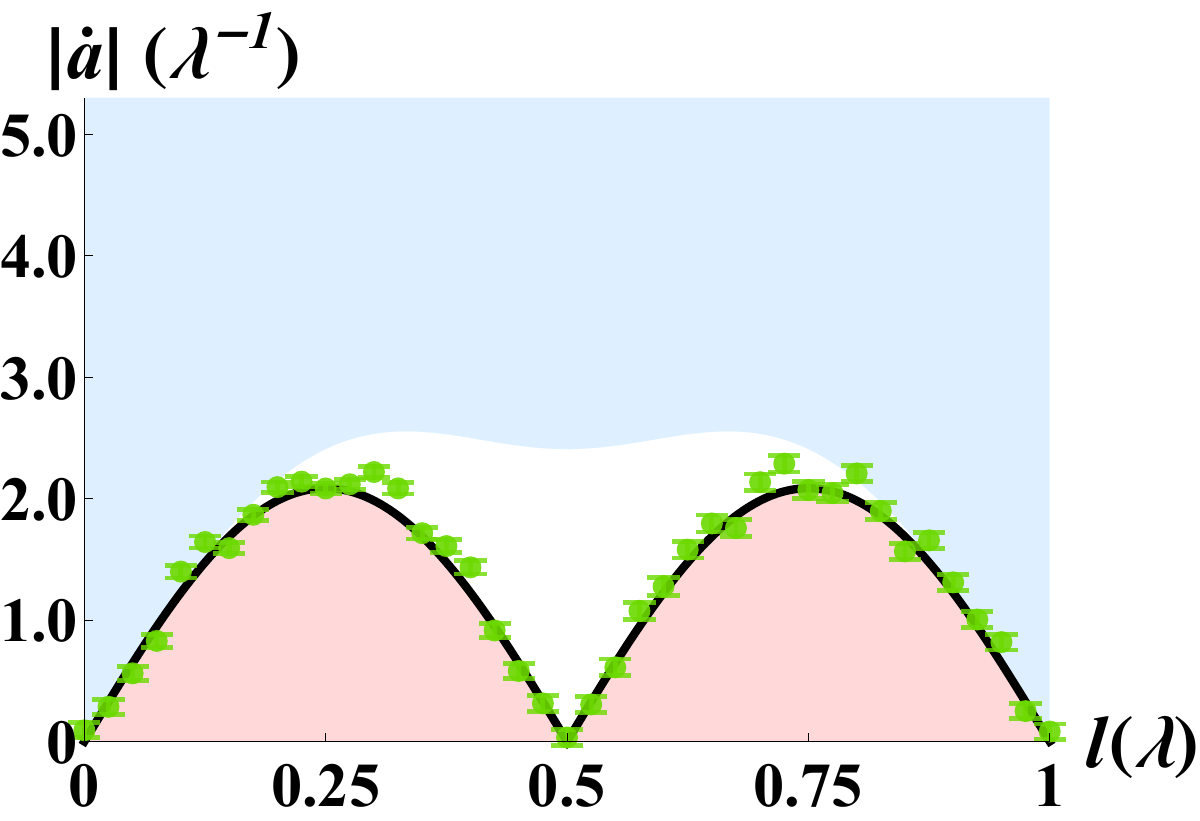}
    \begin{minipage}[t][1in][t]{0.3in}\vspace{0em}(m)\end{minipage}\hspace{-0.1in}\includegraphics[width=1.5in,align=t]{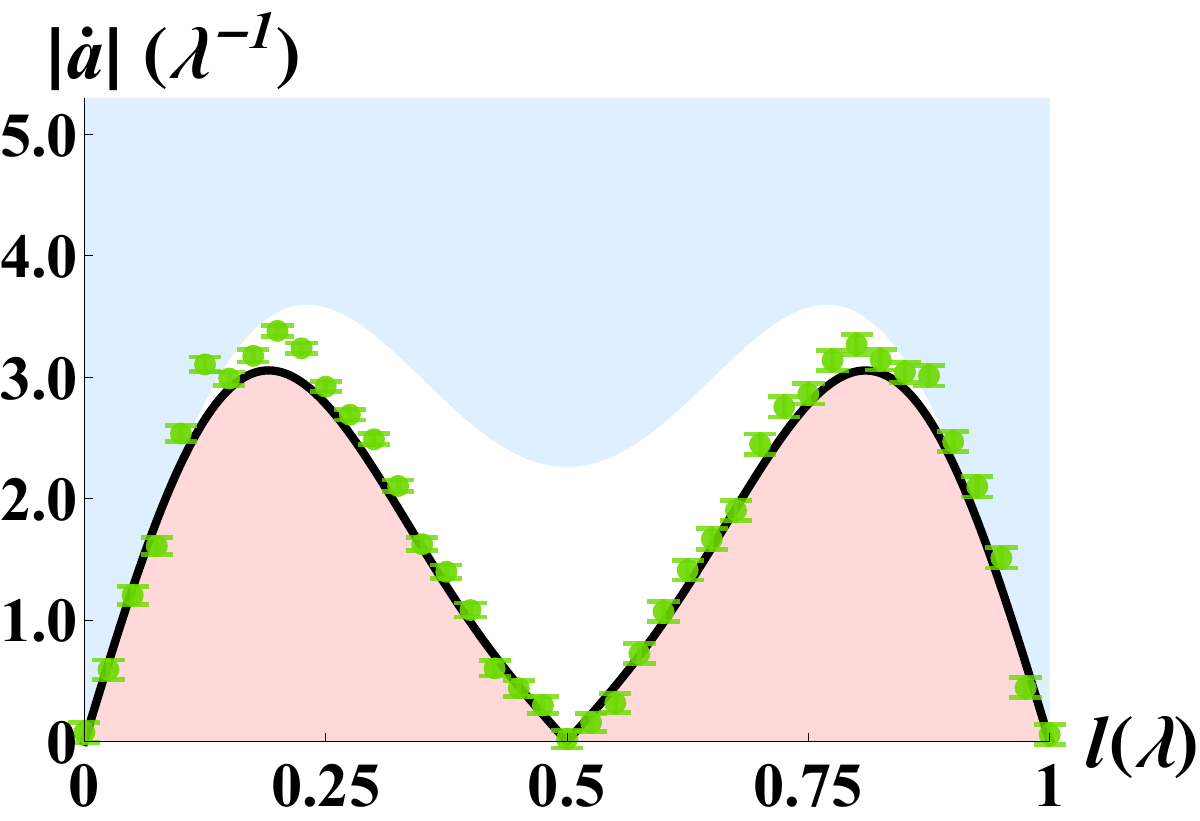}
    \begin{minipage}[t][1in][t]{0.3in}\vspace{0em}(n)\end{minipage}\hspace{-0.1in}\includegraphics[width=1.5in,align=t]{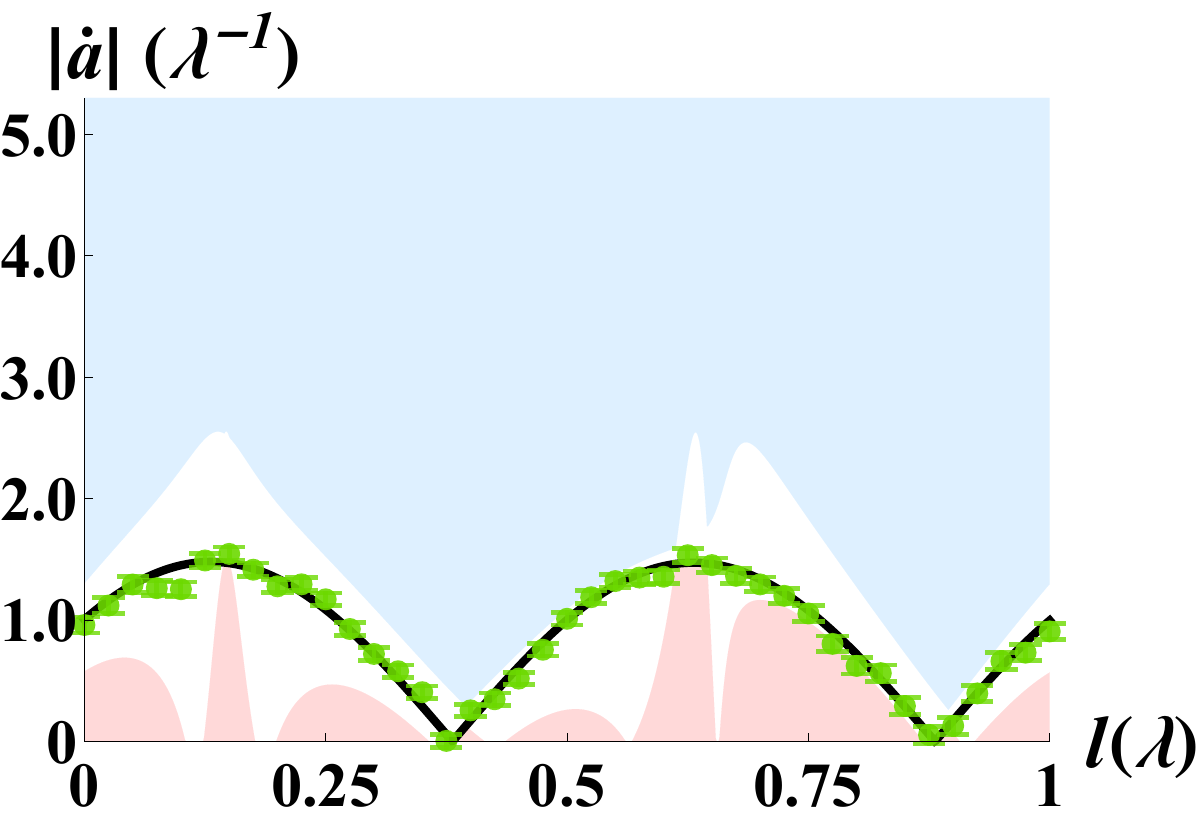}
    \begin{minipage}[t][1in][t]{0.3in}\vspace{0em}(o)\end{minipage}\hspace{-0.1in}\includegraphics[width=1.5in,align=t]{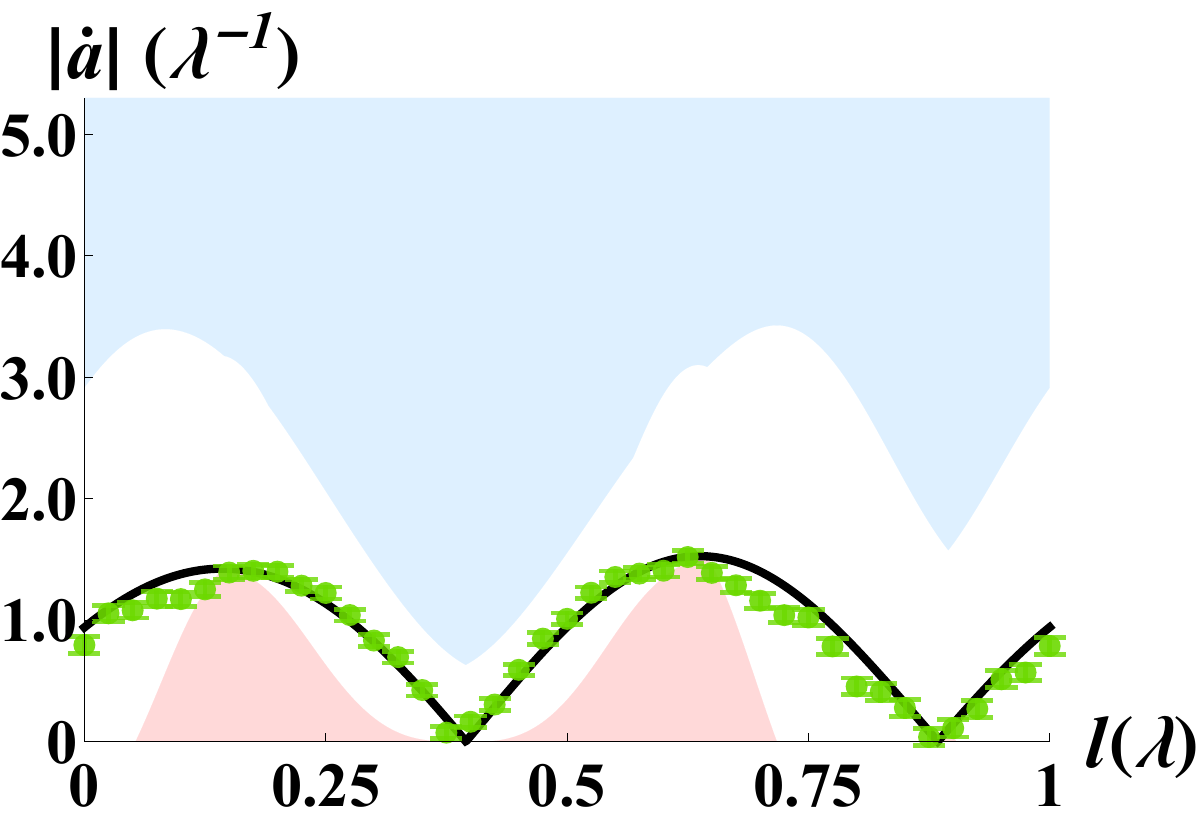}
	\caption{\textbf{Expectation values $\bm{a}$ and quantum speeds $\bm{\left|\dot{a}\right|}$ on observables of the initial state.} \textbf{(a-c, h-k)} represent expectation values $a$. The black line and green points represent the expectation value $a$ in theory and experiments. The error bars are hidden within the experimental points. \textbf{(e-g, l-o)} represent quantum speeds $\left|\dot{a}\right|$ and speed limits on observables of the initial state. The black line and green points represent the quantum speed $\left|\dot{a}\right|$ in theory and experiments, respectively. The blue and red areas indicate the forbidden areas beyond the upper bound and below the lower bound, respectively. Different column pairs except (d) represent different initial states, which are $\qket{+}$, $\qket{++}$, $\qket{\Phi^+}$, $\qket{P}$, $\qket{PP}$, $\qket{P}$, $\qket{PP}$, respectively. Here $\qket{+}=\left(\qket{H}+\qket{V}\right)/\sqrt{2}$, $\qket{\Phi^+}=\left(\qket{HH}+\qket{VV}\right)/\sqrt{2}$, and $\qket{P}$ is a special pure state. There are extra nonunitary noises after the evolution in (j, k, n, o). (d) lists the maximum quantum speeds in (e-g, l-o). According to (e-g) and (l, m), we can see that multiparticles and entanglement can improve quantum speed limits. While nonunitary noise can reduce quantum speed limits according to (l-o). Besides, all points which represent quantum speeds are lying in areas between blue and red areas, which means the tighter quantum speed limits on observables hold no matter in multiparticle or nonunitary systems.}
	\label{label_fig_experiment}
\end{figure}

We experimentally investigate the quantum speed limits of a product two-particle system with an initial state of $\qket{++}$. We scan from $l=0\lambda$ to $l=1\lambda$ and perform tomography at each point to obtain the density matrix $\rho(l)$. The integral time for coincidence counting is 5 seconds. We perform 10,000 Monte Carlo samplings on the measured photon counts and use them to calculate all kinds of physical quantities. We use the average values as the experimental results and the standard deviations as errors. For each sampling, the expectation value and quantum speed are calculated by the central differences:
\begin{align}
    a(l)=&\langle A(l) \rangle=\text{Tr}\left[\rho(l) A\right],\\
    \left|\dot{a}(l)\right|=&\left|\frac{\text{d}a(l)}{\text{d}(l/\lambda)}\right|=\left|\frac{a(l+\Delta l)-a(l-\Delta l)}{2(\Delta l/\lambda)}\right|.
\end{align}
Here, $\Delta l$ represents the step of $l$ in the experiment. Step $\Delta l$ must be an appropriate value according to the specific quantum evolution. Using a too large step will introduce a systematic error, which makes the measured quantum speed lower than the actual quantum speed. Using a too small step will increase the jitter and the standard deviation of the measured quantum speed. We choose the step $\Delta l=0.025\lambda$ for all initial states. With Eq. \eqref{label_eq_upperlower_bound}, we can calculate the tighter quantum speed limits. For a single-particle system with the initial state of $\qket{+}$, the density matrix can be traced from the density matrix with the initial state of $\qket{++}$. As shown in Fig. \ref{label_fig_experiment}, our theoretical predictions accurately match the expectation values $a$ and quantum speed $\left|\dot{a}\right|$ in the experiment. The green points for the quantum speed in the experiment align closely with the black line and consistently remain within the bounds. The maximum upper bound with the $\qket{++}$ initial state, shown in Fig. \ref{label_fig_experiment} (f), is approximately $\sqrt{2}$ times larger than that with the $\qket{+}$ initial state in Fig. \ref{label_fig_experiment} (e); this value corresponds to the standard quantum limit.

For the entangled state, it takes only half the time to rotate to the orthogonal state with respect to a two-qubit product system \cite{MARGOLUS1998188}. Similarly, the upper bound can be up to $\sqrt{2}$ times faster than the product state. We experimentally investigate the tighter quantum speed limits with the initial state of $\qket{\Phi^+}$. The integral time for coincidence counting is 10 seconds. Here, the time to rotate to the orthogonal state is $0.25\lambda$; this value is only half of the $0.5\lambda$ with the $\qket{++}$ or $\qket{+}$ initial state, as shown in Fig. \ref{label_fig_experiment} (a-c). For the maximum quantum speed, as shown in Fig. \ref{label_fig_experiment} (e-g), the fidelity of the entangled state is lower than that of the product state in the experiment, leading to a reduced quantum speed. Maximum quantum speeds are provided in Fig. \ref{label_fig_experiment} (d), and the way to calculate them is shown in supplementary materials \cite{mySupplementalMaterial}. The maximum quantum speed for the $\qket{\Phi^+}$ initial state is $4.981\pm0.011$; this value is higher than $4.049\pm0.012$ for the $\qket{++}$ initial state and within $76.1$ standard deviations, which reflects the Heisenberg limit. The maximum quantum speed for the $\qket{\Phi^+}$ initial state is much higher than $3.103\pm0.006$ for the $\qket{+}$ initial state and within $170.3$ standard deviations.
These results highlight the advantages of entanglement and the many-body system. Note that the entanglement can also reduce the speed limits to almost zero, see in supplementary materials \cite{mySupplementalMaterial}.

Finally, we explore the $\sigma_x$ noisy conditions \cite{Luis2022}. We experimentally investigate the tighter quantum speed limits using the prepared state of $\qket{PP}$. The integral time for coincidence counting is 5 seconds.
As shown in Fig. \ref{label_fig_experiment} (i, m), we use observable $A = \qket{PP}\qbra{PP}$. Then we introduce an $120\lambda$ quartz crystal plate with its optic axis aligned in the $\qket{+}$ direction in each path after the wedge-shaped quartz crystals as $\sigma_x$ nonunitary noise. Unlike other scenarios, when we add $\sigma_x$ noise, as shown in Fig. \ref{label_fig_experiment} (k, o), the lower bound of the quantum speed diverges from the quantum speed itself. However, even in this complex, nonunitary multiparticle open system, tighter quantum speed limits, especially the newly established lower bound, remain valid. 

Using the above evolution data, we can calculate the expectation value $a$ and quantum speed $\left|\dot{a}\right|$ for the initial state of $\qket{P}$ without and with extra noise, as shown in Fig. \ref{label_fig_experiment} (h, j, l, n).
The maximum quantum speed (shown in Fig. \ref{label_fig_experiment} (d)) for the initial state $\qket{PP}$ without extra noise is $3.296\pm0.010$; this value is approximately $1.509$ times larger than $2.184\pm0.004$ for the initial state of $\qket{P}$ without extra noise and shows that the standard quantum limits hold without extra noise in this initial state. However, although extra noise in the Markovian open system provides complex dynamics and mitigates the improvement in the maximum quantum speed caused by the multiparticles, it can increase the quantum speed in some specific small areas of optical path difference $l$, for example, in the area near $l=0$. The maximum quantum speed for the initial state of $\qket{PP}$ with extra noise is $1.391\pm0.004$; this value is nearly equivalent to $1.448\pm0.004$ for the initial state of $\qket{P}$ with extra noise.

\section{Discussion}

In this work, we theoretically discuss multiparticle quantum speed limits on observables and experimentally implement quantum speed limits on observables in single-particle systems and two-particle systems based on an ultrahigh precision and wide-range linear optical platform. First, we introduce tighter quantum speed limits and experimentally investigate the acceleration effect of many-body systems and entanglement. We observe that the maximum quantum speeds for the different initial states of $\qket{+}$, $\qket{++}$, and $\qket{\Phi^+}$ progressively increase. These findings can be applied to the selection of the initial states in quantum metrology, the optimization of quantum battery charging, and the acceleration of quantum computation.
Then, when considering initial states that are far from the eigenstates of the Hamiltonian or noise operator in a Markovian open system, we experimentally demonstrate that the tighter upper bound and newly lower bound of the quantum speed are still established whether in multiparticle or nonunitary open systems, whereas the noise reduces the acceleration effect caused by the increase in the number of particles.

Our work clearly shows that both accelerated and decelerated quantum tasks are possible for multiparticle quantum systems. Therefore, our work paves the way for controlling the quantum speed of
large-scale quantum systems. By addressing the problem of continuously adjusting evolution, we directly show the high-frequency oscillations of the phase components of the density matrix in the quantum evolution. Thus, our results can be used to show the dynamic transient properties in other experiments. Moreover, our study can be theoretically and experimentally extended to more complex quantum systems, for example, the quantum evolution speed limits of non-Markovian and non-Hermitian multiparticle systems, based on the relation between the stochastic operator variance and out-of-time-order correlators \cite{martinez-azconaStochasticOperatorVariance2023}.

\section{Materials and Methods}
\subsection{Experimental Design}

\begin{figure}[H]
	\centering
	\includegraphics[width=7in]{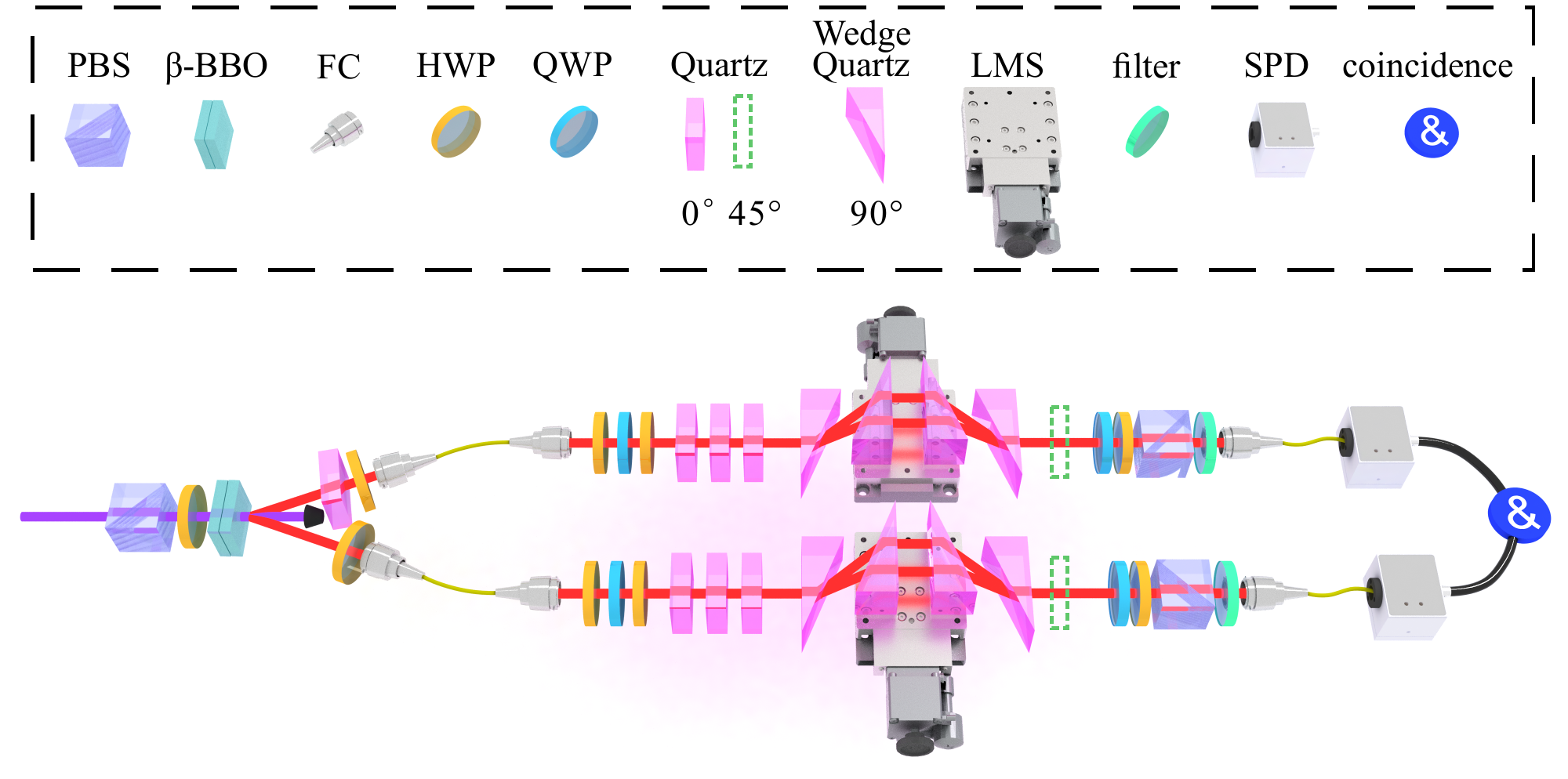}
	\caption{\textbf{Experimental setup for the quantum speed limits implemented by the two-photon dephasing model.} Legend: PBS, polarizing beam splitter; $\beta$-BBO, beta barium borate; FC, fiber coupler; HWP, half-wave plate; QWP, quarter-wave plate; LMS, linear motorized stage; SPD, single photon detector. Entangled photon pairs are produced from $\beta$-BBO. We use filters (at the end of the light path) to control the frequency environment of the open system. QWP and HWP are used to control the polarization system of the open system. The open system evolves into a variable-length quartz crystal system, which can control the evolution time. The optic axes of quartz plates (pink colored) are set to $0^\circ$, while the optic axes of wedge-shaped quartz crystals are set to $90^\circ$. Then each photon may go through an optional quartz plate (green colored) whose optic axis is set to $45^\circ$ for simulating the noise in the nonunitary system. Finally, HWP, QWP, and PBS form the tomography process, and this process can fully extract polarization information within the open system.}
	\label{label_fig_setup}
\end{figure}

A 404 nm continuous-wave laser whose spectral width is approximately 0.06 nm, as shown on the left of Fig. \ref{label_fig_setup}, passes through a polarizing beam splitter (PBS) and a half-wave plate (HWP) to prepare it in the $\qket{V}$ (or $\qket{+}$) polarization state. The beam then passes through two layers of type-I beta barium borate ($\beta$-BBO) crystals with cutting angles of $\theta = 28.89^{\circ}$ positioned perpendicular to each other. The spontaneous parametric down-conversion process in $\beta$-BBO generates idle and signal photons with a central wavelength of 808 nm, as shown in Fig. S1 (a) in \cite{mySupplementalMaterial}. A quartz crystal is placed in the idle path to compensate for the phase differences. Two fiber couplers (FCs) then couple the product state $\qket{HH}$ (or the maximally entangled state $\qket{\Phi^+}$) photon pairs into single-mode fibers. Two HWPs are placed before and after each single-mode fiber to protect the polarization state during transmission.

Both signal and idle photons pass through a quarter-wave plate (QWP) and a HWP to prepare the initial polarization state. Then, they pass through variable-length quartz crystal systems.
We can add or remove $+40\lambda$ quartz crystal plates whose optic axes are aligned in the $\qket{H}$ direction to enable manipulation of $l$ with an accuracy of $40\lambda$. However, an accuracy of approximately $0.025\lambda$ is needed to demonstrate the fine structure during the evolution process. To achieve this, we use four wedge-shaped quartz crystals per path, and their optic axes are aligned in the $\qket{V}$ direction. They have an inclination of $21.8^{\circ}$, and the middle two are mounted on an LMS, allowing horizontal movement to adjust $l$. Four wedge-shaped quartz crystals allow for an $l$ ranging from $-210\lambda$ to $-150\lambda$. The structure and precise optical path of wedge-shaped quartz crystals are shown in the Supplemental Material \cite{mySupplementalMaterial}. By adding five $+40\lambda$ quartz crystal plates, the total $l$ can continuously span from $-10\lambda$ to $50\lambda$.

The QWP-HWP-PBS combination is used to perform tomography measurements on the polarization system. A 12 nm bandpass filter is placed before the FC to block the stray light. Single-photon detectors convert the photon signals into electrical pulses for both photons. A time-to-digital converter is used to capture 3 ns coincidences between the two photons. The coincidence counting rate is approximately 13,000 (or 7200) counts per second for the product state (or maximally entangled state).

We then calibrate the wedge-shaped quartz crystals. We prepare a product state $\qket{++}$ before the quartz crystals and focus on the expectation value on observables $A=\qket{--}\qbra{--}$. Next, we move LMSs to find a position where we obtain the minimum expectation value. This position is referred to as the $l=0\lambda$ point in subsequent steps. Finally, we scan LMSs to find a series of points with local minimum expectation value near the $l=0\lambda$ point and determine that the $102.5\mu\text{m}$ movement results in the $1\lambda$ change in $l$. Since two LMSs have a $0.1\mu\text{m}$ precision, we achieve an accuracy of up to $0.001\lambda$ \cite{mySupplementalMaterial}, which is sufficient for our experiment.

The experimental setup is scalable; thus, we can extend it to more entangled photons easily. The entangled four-photon experiment is used as an example. The entangled two-photon source on the left of Fig. \ref{label_fig_setup} can be replaced by the entangled frequency decorrelated four-photon source \cite{liuzd2024L,Zhong2018,Zhang2015} or entangled frequency correlated four-photon source \cite{Hubel2010,Hamel2014,Zhao2025,Kopf2025}. The right top part and the right bottom part of Fig. \ref{label_fig_setup} are the same. So, we can build them four times to receive photons from the four-photon source. The tomography measurements still need only one QWP, one HWP, and one PBS per path, and the only concession is additional measurement bases and a multichannel time-to-digital converter. In summary, both the theory and the experiment can be changed to a multiparticle version. Even if we consider a certain range of fidelity decline, as the number of particles $N$ increases, the acceleration effect of entangled states will become more notable \cite{mySupplementalMaterial}.

%\appendix

% \nocite{*}

\bibliographystyle{apsrev4-2}
\bibliography{speedlimit}% Produces the bibliography via BibTeX.

\section{Acknowledgments}

% \begin{acknowledgments}
\subsection{Funding:}
This work was supported by the Innovation Program for Quantum Science and Technology (No. 2021ZD0301200), the National Natural Science Foundation of China (No. 11821404). Z. -D. Liu was supported, in part, by the Anhui Provincial Natural Science Foundation (No. 2408085MA010).
% \end{acknowledgments}

\subsection{Author contributions:}

C. -F. L., Z. -D. L., and R. -H. M. planned and designed the experiments. R. -H. M. and Z. -D. L. implemented the experiments with the help of C. -X. N., Y. -C. H. and H. Z. under the supervision of C. -F. L. and G. -C. G. The theoretical analysis was performed by R. -H. M. The paper was written by R. -H. M. and Z. -D. L. All authors discussed the contents.

\subsection{Competing interests:}

Authors declare that they have no competing interests.

\subsection{Data and materials availability:}

All data needed to evaluate the conclusions in the paper are present in the paper and/or the Supplementary Materials.

\section{Figures and Tables}

\renewcommand{\arraystretch}{1.5}%表格行高倍率

\end{document}